\documentclass[conference]{IEEEtran}

\usepackage{booktabs}
\usepackage{cite}
\usepackage{multirow}
\usepackage{microtype}
\usepackage{xspace}
\usepackage{amsmath}
\usepackage{amssymb}
\usepackage{amsthm}
\usepackage{algorithm}
\usepackage{algorithmic}
\usepackage{graphicx}
\usepackage{url}
\usepackage[hidelinks]{hyperref}
\usepackage{needspace}

\newtheorem{theorem}{Theorem}

\newcommand{\sysname}{EnvelopeTC\xspace}
\newcommand{\pc}{PolicyCache\xspace}
\newcommand{\hatmodel}{HAT\xspace}

\newcommand{\Description}[1]{}

\begin{document}
\raggedbottom

\title{\sysname: Hierarchical Intra-Path Learning for Adaptive Traffic Control}

\author{
\IEEEauthorblockN{Wenyang Jia$^{1}$,
                  Jingjing Wang$^{1}$,
                  Ziwei Yan$^{1}$,
                  Tanren Liu$^{2}$,
                  Yakun Ren$^{2}$,
                  Kai Lei$^{1,\dagger}$}
\IEEEauthorblockA{$^{1}$ICNLab, Shenzhen Graduate School, Peking University,
                  Shenzhen, P.R.China}
\IEEEauthorblockA{$^{2}$SF Technology, Shenzhen, P.R.China}
\IEEEauthorblockA{$^{\dagger}$Corresponding author: leik@pkusz.edu.cn}
}

\maketitle

\begin{abstract}
Software defined networks offer global visibility, yet centralized control loops are too slow for transient congestion and bursty traffic dynamics. Existing learned traffic control schemes often rely on offline training, making them fragile under distribution shifts. We present \sysname{}, a hierarchical SDN traffic control framework that enables local online adaptation under centralized policy control. Its key abstraction is a policy envelope: the controller compiles network wide intent into bounded per path action spaces, while edge agents learn and execute metering, queueing, and rerouting decisions only within those bounds. Policy envelopes also make local actions auditable and reversible when they affect shared bottlenecks. Evaluation on a 1,024 host software SDN testbed shows that \sysname{} improves average core link utilization by 35.5\% over Static ECMP and 18.3\% over Centralized TE. It reduces elephant flow P99 FCT by 34.3\% over end host congestion control, lowers SLA violations from 18.2\% to 6.8\%, and uses less than 2\% CPU and 12\,MB memory per edge agent.

\noindent 
The source code is available in an anonymized repository at \url{https://anonymous.4open.science/r/JCC2026-EnvelopeTC/}.
\end{abstract}

\begin{IEEEkeywords}
software-defined networking, online learning, traffic engineering,
congestion control, intra-path learning, Hoeffding Adaptive Tree
\end{IEEEkeywords}

\section{Introduction}
\label{sec:intro}

Software-defined networking has become the dominant paradigm for operating large-scale
data-center and wide-area networks. By centralizing control in a programmable controller,
SDN enables fine-grained traffic-engineering policies, tenant isolation, and network-wide
topology responses from a single vantage point~\cite{mckeown2008openflow,kim2013improving}.

This global visibility, however, creates a fundamental tension with fast traffic control.
Congestion events, elephant-flow bursts, and link microbursts evolve on timescales of tens
of milliseconds, comparable to a single RTT. Routing per-RTT control decisions through a
centralized controller incurs at least one controller round-trip plus processing delay,
making purely centralized fast-loop control impractical at
scale~\cite{curtis2011devoflow}.

The networking community has responded with two broad strategies. \emph{Static policy}
programs switch pipelines with pre-computed ECMP weights, meter rules, or queue
assignments; it is fast but inflexible, as it cannot adapt to dynamic traffic matrices or
link-quality changes. \emph{Offline learning} trains a neural network or RL agent on
historical traces and deploys the resulting policy to the controller or data plane; this
approach adapts within the training distribution but fails to generalize when real traffic
deviates from training conditions, a well-documented limitation of inter-flow
learning~\cite{jay2019internet}.

This paper presents \sysname{}, a hierarchical SDN traffic-control architecture based on
\emph{intra-path learning}: edge agents learn and execute fast local actions while the
controller manages global intent, action bounds, and coordination. The contribution is not
a new learning model; it is a control abstraction that makes locality-based online learning
safe and composable for SDN actions whose effects extend beyond the local learning scope.

\sysname{} makes three contributions:

\begin{itemize}
  \item \textbf{Intra-path learning.} We scope each policy cache to a path, bottleneck,
    or tenant aggregate; training and execution are confined to that aggregate only.

  \item \textbf{Policy envelopes.} The controller computes rate bounds, reroute
    permissions, and utility weights; agents enforce these envelopes before issuing any
    switch action.

  \item \textbf{Safe multi-agent coordination.} Bottleneck alerts, action logs, and
    primary-agent arbitration serialize conflicts. We implement \sysname{} with Ryu,
    Open vSwitch, and gRPC, and evaluate it against nine baselines on a 1,024-host testbed.
\end{itemize}
\section{Background and Motivation}
\label{sec:background}

\subsection{SDN Traffic Engineering Today}

Modern SDN deployments address traffic engineering at two timescales: the \emph{slow
timescale} (seconds to minutes), where the controller installs or updates forwarding rules
in response to topology events and traffic-matrix shifts; and the \emph{fast timescale}
(sub-second), where individual flows encounter congestion requiring millisecond-level rate
adjustment, rerouting, or queue management.

Existing mechanisms handle fast-timescale control poorly. Static OpenFlow meter rules
apply fixed rate caps regardless of load; ECMP uses fixed weights that cannot react to
transient hot spots; and threshold-based rerouting triggers only above a static utilization
threshold, causing oscillation under dynamic traffic.

\subsection{The Generalization Problem in SDN}

Several works have proposed replacing hand-crafted SDN policies with learned ones. Offline
reinforcement learning~\cite{chinchali2018cellular} trains an agent on
simulated or historical traces and deploys it to the controller. These approaches share a
common limitation: the fixed policy degrades when real conditions differ from the training
distribution (the \emph{inter-flow generalization} problem~\cite{tian2026policycache}).
Deploying an Aurora-architecture agent~\cite{jay2019internet} under shifted traffic
matrices or link-failure hot spots degrades utility by up to 40\%, consistent with this
limitation.
\subsection{From Intra-Flow to Intra-Path Learning}

\pc~\cite{tian2026policycache} proposes \emph{intra-flow learning}: training and execution
are confined to a single flow, avoiding cross-flow generalization. \sysname{} reuses this
locality principle and the same non-parametric \hatmodel{} family, lifting the scope to an
SDN \emph{path aggregate}: an active path, bottleneck, tenant, or service class observed
and controlled by one edge agent. The learned policy applies only to that aggregate and is
invalidated when path conditions change. This lift introduces systems challenges absent in
single-flow CC: rerouting can displace congestion to another link, meter updates can affect
co-located tenants, and two agents on the same bottleneck can oscillate. \sysname{}
addresses these through controller-compiled policy envelopes, shared-resource monitoring,
and conflict arbitration.

\section{\sysname{} Architecture}
\label{sec:arch}

\begin{figure}[t]
  \centering
  \includegraphics[width=\columnwidth]{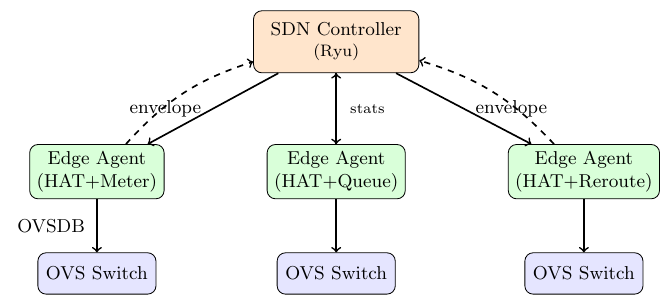}
  \caption{\sysname{} two-plane architecture. The SDN controller manages global state and
           distributes policy envelopes; edge agents run intra-path learning and execute
           fast SDN actions.}
  \Description{Two-plane architecture with a central SDN controller distributing policy
               envelopes to multiple edge agents, which observe local telemetry and execute
               switch actions.}
  \label{fig:arch}
\end{figure}

\sysname{} consists of two control planes operating at different timescales and
granularities (Figure~\ref{fig:arch}).

\subsection{Central SDN Controller: Slow Global Control}

The SDN controller operates at the \emph{slow timescale} (hundreds of milliseconds to
seconds), maintaining the topology graph and backup paths, aggregating per-link telemetry
from sFlow/INT streams, computing policy envelopes for active aggregates, detecting and
arbitrating conflicting agent actions at shared bottlenecks, and monitoring agent behavior
with override and rollback capability.

\textbf{Envelope computation.}
The controller computes envelopes from measured demand, topology, tenant policy, and
service templates. For each congested link $b$ with capacity $C_b$ and active aggregates
$A_b$, it reserves tenant floors $m_i$ and headroom $\epsilon_q C_b$; the remaining budget
$H_b=(1-\epsilon_q)C_b-\sum_{j\in A_b}m_j$ is allocated by weighted water filling:
\[
\begin{aligned}
r_{\min,i} &= m_i,\\
r_{\max,i} &= \min\!\left\{\hat d_i(1+\rho),\; R_i^{\mathrm{tenant}},\right.\\
&\qquad\left. m_i+H_b\frac{\phi_i}{\sum_{j\in A_b}\phi_j}\right\}.
\end{aligned}
\]
For a path crossing multiple constrained links, the agent receives the tightest per-link
bound, making fairness and tenant isolation controller-enforced: an agent may learn how to
use its allocation but cannot exceed tenant ceilings or reroute outside permitted paths.

\textbf{Utility weights.}
The utility vector $\mathbf{w}$ is initialized from service-class templates and adjusted at
each refresh: sustained SLA violations increase $w_\mathrm{lat}$ and $w_\mathrm{sla}$;
spare capacity with satisfied SLAs increases $w_\mathrm{thr}$. Weights are normalized and
clipped to operator ranges, so utility adaptation changes the local objective without
granting new action authority.

\noindent A policy envelope is the contract that constrains exploration, clips model
predictions before execution, provides a versioned audit object, and enables safe fallback
when stale. The controller does \emph{not} make per-RTT decisions; it bounds which
per-interval decisions are legal.

\subsection{Edge Agent: Fast Local Online Learning}

Each \emph{edge agent} is deployed at a ToR switch CPU, vSwitch, P4 control plane, or
SmartNIC. It maintains one \emph{policy cache} per monitored path or traffic aggregate,
implemented as a Hoeffding Adaptive Tree (\hatmodel{}) as in
\pc~\cite{tian2026policycache}.

The agent operates in two modes. In \textbf{backup exploration mode}, when the cache is
untrained or underperforming ($p < p_\mathrm{th}$), it probes envelope-valid candidate
actions on canary traffic and uses the higher-utility action as the empirical label. In
\textbf{model execution mode}, once $p \geq p_\mathrm{th}$, the \hatmodel{} predicts an
action, the enforcer clips it to the envelope, reroutes are shadow-checked, and the
committed action is logged for audit and rollback. If the envelope becomes stale, agents
disable new reroutes and keep only meter/queue actions inside the last valid bounds.
ADWIN~\cite{bifet2007learning} triggers a return to exploration when path conditions shift.

\subsection{Controller--Agent Interface}

The interface is lightweight and asynchronous: the controller pushes policy envelopes,
bottleneck alerts, and overrides, while agents return telemetry, model state, and action
logs. Envelopes carry a version and staleness bound; agents apply updates on the next
action cycle without requiring flow-level rule refreshes. This is what distinguishes
\sysname{} from a plain intra-flow learner on an SDN switch: local learning is
permitted only inside a controller-compiled action set.

\section{Learning Formulation and Analysis}
\label{sec:learning}

\subsection{State}

The state vector observed by an edge agent at each measurement interval $t$ is:
\[
\begin{aligned}
\mathbf{s}_t = [&\mathrm{util}_t,\; q_t,\; \mathrm{loss}_t,\; \mathrm{ecn}_t,\;
                 \mathrm{thr}_t,\; \mathrm{delay}_t,\\
               &\Delta\mathrm{util}_{t-1},\; \Delta\mathrm{thr}_{t-1},\;
                 \Delta\mathrm{delay}_{t-1},\; a_{t-1},\; a_{t-2}]
\end{aligned}
\]
where $\Delta(\cdot)$ denotes the one-interval relative change.
Relative-change features capture temporal dynamics without requiring cross-path normalization,
consistent with the intra-path learning philosophy.

\subsection{Action}

\sysname{} controls SDN-level actions rather than TCP~cwnd. The action space is discrete
and path-specific; at each interval the agent selects one action per dimension:
\textbf{meter-rate adjustment} (increase or decrease by step $\alpha$, clipped to
$[r_\mathrm{min}, r_\mathrm{max}]$);
\textbf{queue assignment} (promote or demote the flow class by one priority level); and
\textbf{elephant-flow rerouting} (binary trigger/release for flows exceeding 10\,MB in the
measurement window).
ECN marks are used as telemetry features; the prototype does not tune ECN marking thresholds.

\subsection{Utility Function}

\begin{equation}
\begin{aligned}
u_t ={}& w_\mathrm{thr}\,\Delta\mathrm{thr}_t
       - w_\mathrm{lat}\,\Delta\mathrm{delay}_t
       - w_\mathrm{loss}\,\mathrm{loss}_t \\
      &- w_\mathrm{sla}\,v_t
       - w_\mathrm{act}\,|a_t - a_{t-1}|
\end{aligned}
\label{eq:utility}
\end{equation}

\noindent where $v_t$ is the SLA violation indicator.
The controller distributes a utility vector
$\mathbf{w}=(w_\mathrm{thr}, w_\mathrm{lat}, w_\mathrm{loss}, w_\mathrm{sla}, w_\mathrm{act})$
as part of the policy envelope, using the service-class templates and bounded adjustment
rules described in Section~\ref{sec:arch}.

\subsection{Empirical Labels}

\sysname{} does not assume access to a global optimal action. During backup exploration,
the agent obtains an \emph{empirical label} $a^\dagger(\mathbf{s})$ by comparing two
envelope-valid candidate actions over adjacent measurement intervals and selecting the one
with higher observed utility. Thus $a^\dagger$ is a local, path-specific target, not a
network-wide optimum, and remains valid only while the local distribution, envelope, and
neighboring agents' actions are sufficiently stable over the exploration window.

\subsection{Envelope Refresh and Guarantees}

At each slow-timescale refresh, the controller initializes each aggregate's floor and cap
from tenant policy, tightens $r_{\max}$ on congested links by weighted water filling,
sets the reroute permission bit from residual capacity and cooldown state, and pushes the
versioned envelope $\langle r_{\min},r_{\max},\pi_{\mathrm{rt}},\mathbf{w}\rangle$ to the
responsible agent. Safety mechanisms are procedural: exploration is canaried, actions are
envelope-clipped, reroutes are shadow-checked, harmful actions trigger rollback, and $p$
is tracked as a moving average of prediction accuracy.

\subsection{Analysis Scope and Assumptions}

The results below are conditional and operational; they characterize predictable behavior,
not global optimality. We assume: (A1)~within each exploration window, each visited state
region has stable expected utility with bounded measurement noise; (A2)~the stream visits a
finite set $S$ of regions often enough for the cache to receive repeated samples; (A3)~canary
and micro-exploration continue throughout; (A4)~agents enforce the latest envelope and the
controller serializes conflicting meter-rate updates.

\begin{theorem}[Single-Agent Cache Consistency]
\label{thm:single}
Under A1--A3, after collecting
$N_\mathrm{conv} = O\!\left(\frac{R^2}{\varepsilon^2}\ln\frac{|S|}{\delta}\right)$
samples, the \hatmodel{} predicts $a^\dagger(\mathbf{s})$ for all visited regions with
probability at least $1-\delta-\eta$ ($\eta$ bounds exploration-noise mislabeling).
Once accuracy exceeds $p_\mathrm{th}$ by a margin, the agent switches to model execution
after $O(1/(1-\beta))$ additional samples.
\end{theorem}

\begin{theorem}[Policy-Envelope Compliance]
\label{thm:safety}
Every action executed by an \sysname{} agent satisfies
$\langle r_\mathrm{min}, r_\mathrm{max}, \pi_\mathrm{reroute}\rangle$, provided the agent
has received the envelope and applies the enforcer before OVSDB execution. This is a
syntactic guarantee and does not imply SLA or network-wide invariants under stale
telemetry or conflicting reroutes.
\end{theorem}

\begin{theorem}[Conditional Recovery after Drift]
If path conditions change at $t_d$ and restabilize, and the \hatmodel{} error rate rises
from $p_1$ to $p_2 > p_1$, ADWIN detects the drift within
$T_\mathrm{detect} = O\!\left((p_2-p_1)^{-2}\right)$ samples (w.h.p.). The agent then
recovers cache consistency after $N_\mathrm{conv}$ additional labeled samples under A1--A3.
\end{theorem}

\section{Implementation}
\label{sec:impl}

\sysname{} is implemented as three software components.

\textbf{SDN Controller (Ryu + \sysname{} Module).}
We extend Ryu~\cite{ryu} with an \sysname{} controller module for global monitoring, policy
envelope generation, bottleneck detection, and multi-agent coordination. The module
subscribes to sFlow records from OVS instances and maintains per-path envelope state in an
in-memory store.

\textbf{Edge Agent (Python Daemon).}
Each agent runs as a Python daemon co-located with the vSwitch control plane. It provides:
(i)~a telemetry collector polling OVS port counters and queue statistics via OVSDB every
50\,ms; (ii)~a \hatmodel{} model per monitored path using the River ML
library~\cite{river}; (iii)~an action executor translating model outputs to
meter/queue/flow-table updates via OVSDB RPC; and (iv)~a policy-envelope enforcer clipping
all actions to controller-provided bounds. One-way delay is measured from timestamped
endpoint probes, not inferred from OVSDB counters.

\textbf{Controller--Agent Transport (gRPC).}
Policy envelopes and reports are exchanged over gRPC. Envelopes are pushed on change;
agents apply updates within a 500\,ms staleness bound.

\textbf{Action dimensions.}
The prototype implements three action dimensions: (1)~OpenFlow meter rate adjustment
($\alpha = 10\%$, step size), (2)~OVS queue assignment across three priority levels, and
(3)~elephant-flow rerouting via priority flow-table rules.

\section{Evaluation}
\label{sec:eval}

\noindent We evaluate \sysname{} on overall performance, tail latency, overhead, and
robustness under controller and telemetry stress.

\subsection{Methodology and Baselines}

Experiments run on 1,024 AWS c5.xlarge instances emulating a 64-rack Clos fabric with OVS
bridges, GRE tunnels, 10\,Gbps \texttt{tc}-shaped logical links, and OpenFlow group-table
routing. A Ryu controller on a c5.2xlarge manages 64 rack agents over gRPC; agents poll
OVS counters every 50\,ms and enforce a 500\,ms reroute cooldown. Long flows use iperf3;
short flows use fixed-seed replay across rack pairs. We report 10 trials (seeds
1000--1009) with 95\% Student-$t$ confidence intervals; FCT comes from application logs and
one-way delay from 100\,Hz timestamped probes. Because this is a software public-cloud
fabric, absolute latency and reordering include VM, GRE, and OVSDB noise and should be
read as relative comparisons.

All schemes share the same topology, traces, and measurement pipeline. Baselines are
Static ECMP, Centralized TE with a 500\,ms loop, Threshold Reroute above 80\% utilization,
Static Meter, Heuristic Queue, frozen Aurora-SDN~\cite{jay2019internet}, LetFlow-style
flowlets~\cite{vanini2017letflow}, Presto-style flowcells~\cite{he2015presto}, and
\pc{} end-host congestion control~\cite{tian2026policycache}. Aurora-SDN is trained
offline for $10^6$ steps and evaluated without online fine-tuning; claims against learned
baselines are limited to this frozen setting and to \pc{}'s end-host learner.

We use three workloads: \textbf{elephant-heavy} (80\% of bytes from flows $>$100\,MB;
matrix shifts every 60\,s), \textbf{mice-heavy} (Pareto mean 50\,KB, $\kappa{=}1.2$), and
\textbf{mixed real-time} (40\% latency-sensitive flows with a 10\,ms one-way SLA).

\subsection{Link Utilization}

\begin{figure}[t]
  \includegraphics[width=\columnwidth]{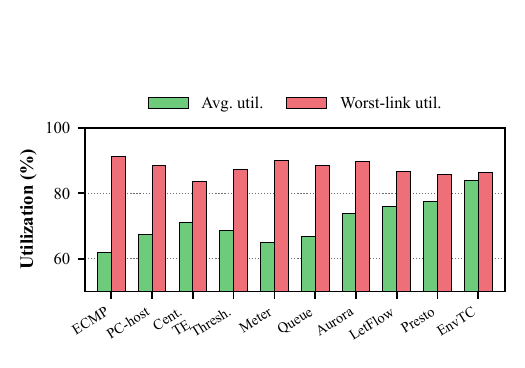}
  \caption{Link utilization: average vs.\ worst-link maximum across all schemes.
           \sysname{} achieves the highest average utilization with a small max-average gap.}
  \Description{Grouped bar chart comparing average utilization and maximum core-link
               utilization for all baseline schemes and EnvelopeTC.}
  \label{fig:util}
\end{figure}

Figure~\ref{fig:util} shows average and worst-link utilization under the elephant-heavy workload.

\sysname{} achieves \textbf{35.5\%} higher average utilization than Static ECMP
(84.0\% vs.\ 62.0\%) and \textbf{18.3\%} higher than Centralized TE. Its worst-link
maximum is 86.4\%, only 2.4\,pp above its average. \pc{} (end-host) yields only 67.4\%:
end-host CC cannot see bottleneck locations or reroute elephant flows, confirming that
network-level intra-path learning provides capabilities beyond end-host control.
Aurora-SDN performs well within its training distribution but degrades by 6.8\,pp under
shifted traffic matrices. LetFlow-style and Presto-style improve over Static ECMP but
remain below \sysname{} under sustained hot spots.

\subsection{Flow Completion Time}

\begin{figure}[t]
  \centering
  \begin{minipage}[b]{0.49\columnwidth}
    \includegraphics[width=\linewidth]{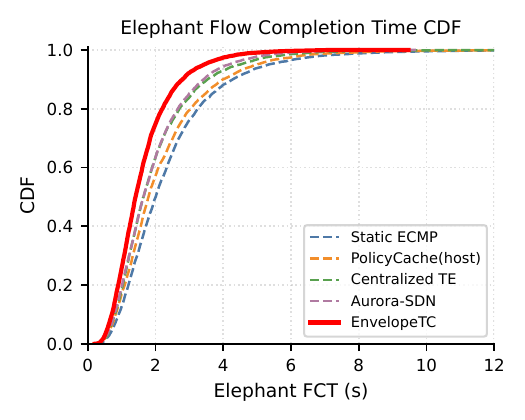}
    \centerline{\footnotesize (a) Elephant FCT ($>$10\,MB)}
  \end{minipage}%
  \hfill
  \begin{minipage}[b]{0.49\columnwidth}
    \includegraphics[width=\linewidth]{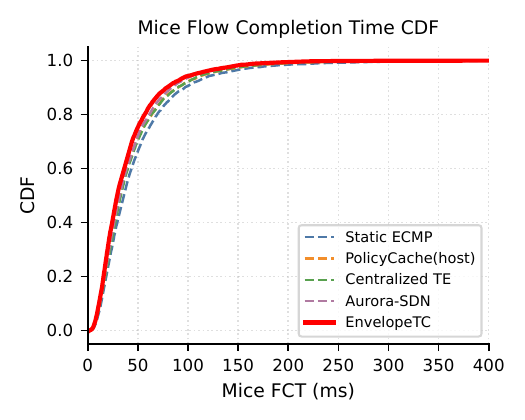}
    \centerline{\footnotesize (b) Mice FCT ($<$100\,KB)}
  \end{minipage}
  \caption{Flow Completion Time CDFs (elephant-heavy workload). \sysname{} shifts the
           elephant CDF leftward by 33\% vs.\ Static ECMP while also improving mice FCT.}
  \Description{Two CDF plots showing elephant and mice flow completion times; EnvelopeTC
               has the leftmost curves among compared schemes.}
  \label{fig:fct}
\end{figure}

\sysname{} reduces elephant mean FCT by \textbf{33.2\%} and P99 FCT by \textbf{40.3\%}
relative to Static ECMP. The improvement over Centralized TE (18.7\% mean, 24.1\% P99)
stems from faster reaction: edge agents detect queue buildup within the current 50\,ms
interval and execute reroutes within the same RTT, whereas Centralized TE requires a full
sFlow aggregation cycle ($\approx$500\,ms). \pc{} (end-host) reduces elephant mean FCT by only 9.5\% relative to Static ECMP because
it cannot reroute elephant flows, a fundamental limitation of end-host CC.

\subsection{Tail Latency and SLA Compliance}

\begin{figure}[t]
  \centering
  \includegraphics[width=\columnwidth]{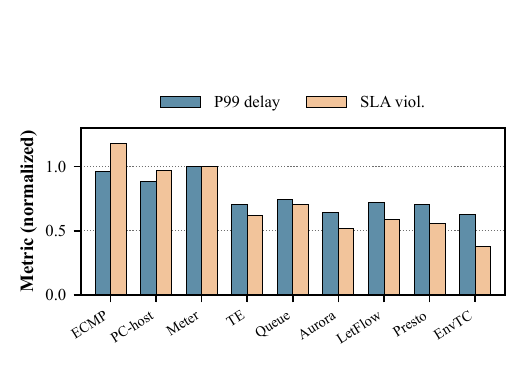}
  \caption{Tail latency and SLA compliance under the mixed real-time workload, normalized
           to Static Meter. \sysname{} has both the lowest P99 delay and the lowest
           violation rate.}
  \Description{Grouped bar chart comparing normalized P99 one-way delay and SLA violation
               rate for real-time flows across evaluated schemes.}
  \label{fig:sla_summary}
\end{figure}

\sysname{} reduces P99 delay for real-time flows by \textbf{37.7\%} compared to Static
Meter and by 11.7\% compared to Centralized TE. SLA violation rate drops from 18.2\%
(Static Meter) to \textbf{6.8\%}, a 62.6\% relative reduction. \pc{} (end-host) achieves
17.6\% violation rate; it reduces P99 delay slightly by backing off cwnd but cannot
promote flows to higher-priority queues, a capability unique to SDN-level control. Aurora-SDN achieves 9.4\% within its training distribution but rises to 13.1\%
under out-of-distribution traffic, consistent with the limitations of frozen offline
policies in this setup.

\subsection{Convergence, Overhead, and Coordination}

\begin{figure}[t]
  \centering
  \includegraphics[width=\columnwidth]{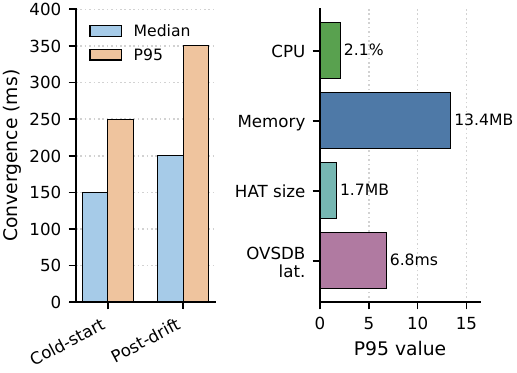}
  \caption{Agent convergence and overhead summary. Convergence remains below 350\,ms at P95,
           with small per-agent CPU, memory, model-size, and OVSDB-latency overheads.}
  \Description{Two-panel grouped bar chart showing cold-start and post-drift convergence
               medians and P95s, plus P95 CPU, memory, HAT model size, and OVSDB
               action latency values.}
  \label{fig:overhead}
\end{figure}

All agents reach model execution mode within 400\,ms from cold-start
(Figure~\ref{fig:overhead}). Agent CPU stays below 2.1\% and memory below 13.4\,MB.
Without controller arbitration, agents sharing a bottleneck oscillate at 4.3 reroute-flip
events/s with $\pm$22\,pp utilization variance; with arbitration, flips drop to 0.4/s and
variance narrows to $\pm$5.1\,pp, adding $<$0.3\% to controller--agent traffic.

\subsection{Ablation and Sensitivity}

We ablate each action type on the elephant-heavy workload.
Rerouting is most impactful: disabling it reduces average utilization from 84.0\% to
76.3\% ($-$7.7\,pp) and elephant mean FCT from 1.61\,s to 2.09\,s. Queue-priority
contributes $+$3.9\,pp and metering $+$1.9\,pp. \sysname{} is robust across parameter
settings: at $\alpha{=}10\%$ (default), utilization gain over Static Meter is 19.1\,pp;
$\alpha{=}2\%$ reduces it to 6.1\,pp, and $\alpha{=}25\%$ speeds convergence but adds
2.4\,pp SLA violations. Coarsening the interval from 50\,ms to 200\,ms costs 4.7\% in
utilization and 8.2\% in P99 FCT. Under shifted traffic matrices, ADWIN-triggered
re-exploration converges within 200--350\,ms.

\subsection{Controller Loop and Robustness}

\textbf{Centralized TE update interval.}
TE-loop sensitivity shows that faster loops improve utilization
but increase CPU and rule churn sharply. Even a 50\,ms loop remains below \sysname{}:
global collect--optimize--install cannot match our design's split 50\,ms agent /
500\,ms envelope cadence.

\textbf{Stability and side effects.}
Stress tests cover stale envelopes, telemetry loss, slow OVSDB
commits, controller outage, rapid traffic oscillation, and excessive reroute frequency.
During a controller outage, agents continue meter and queue actions inside the last valid
envelope but disable new reroutes after the 500\,ms staleness bound. Disabling the reroute
cooldown improves neither utilization nor FCT but increases flips and reordering.

\subsection{Discussion and Limitations}

The results suggest that \sysname{} is most useful when the controller can express
operator intent as narrow action bounds and local agents can observe enough short-term
telemetry to choose among those legal actions. This matches software fabrics, virtualized
edge deployments, and SmartNIC-assisted designs where meter, queue, and selective reroute
actions are already available to the control plane. The design is less appropriate when
the data plane exposes only static ECMP weights or when telemetry arrives slower than the
congestion episode being controlled.

In deployment, envelope width is the main operator knob. Narrow envelopes reduce risk and
make rollback easier, while wider envelopes allow faster recovery from traffic-matrix
shifts but require stricter audit and arbitration. This separation lets operators tune
safety policy at the controller without retraining the edge learner.

Our evaluation intentionally separates relative system behavior from hardware claims.
The OVS/GRE testbed exercises controller-agent timing, online adaptation, reroute
coordination, and failure handling at scale, but it does not reproduce ASIC queue dynamics,
lossless RDMA fabrics, or in-switch congestion feedback. We therefore view hardware P4 or
SmartNIC validation as a deployment step rather than as evidence needed for the core
control abstraction. A second limitation is baseline scope: Aurora-SDN is a frozen
offline-RL baseline, so the comparison tests robustness to distribution shift, not the
full design space of online-adapting multi-agent RL.

\section{Related Work}
\label{sec:related}

\textbf{SDN traffic engineering and load balancing.}
B4~\cite{jain2013b4}, SWAN~\cite{hong2013achieving}, and
Hedera~\cite{alfares2010hedera} optimize routing at controller timescale, while
CONGA~\cite{alizadeh2014conga}, HULA~\cite{katta2016hula}, LetFlow~\cite{vanini2017letflow},
and Presto~\cite{he2015presto} address ECMP imbalance through in-network feedback,
flowlets, or flowcells. \sysname{} is complementary: it keeps global policy at the
controller but gives edge agents bounded authority over SDN meters, queues, and selective
reroutes.

\textbf{Edge admission and defense.}
Edge-side mechanisms can also adapt ingress behavior for security: EdgePoW~\cite{jia2026edgepow}
uses ingress-aware non-interactive proof-of-work to defend against volumetric SYN floods.
\sysname{} targets a complementary control problem, using bounded edge authority for
adaptive traffic control rather than attack admission.

\textbf{Online learning and hierarchical control.}
Vivace~\cite{dong2018pcc}, Proteus~\cite{du2021unified}, and
\pc~\cite{tian2026policycache} show that online exploration can adapt transport behavior
inside a local scope. \sysname{} lifts that locality to SDN path aggregates, where actions
can affect tenants and neighboring paths; policy envelopes and arbitration make such local
learning auditable. ONIX~\cite{koponen2010onix} and Kandoo~\cite{hassas2012kandoo}
similarly layer control for scalability, but their local elements process events rather
than execute bounded online learning policies.

\section{Conclusion}
\label{sec:conclusion}

We presented \sysname{}, a hierarchical SDN traffic-control framework that lifts
intra-flow locality from end-host congestion control to SDN traffic aggregates. The
main systems contribution is not a new tree learner, but the controller--agent abstraction
needed to make local online learning composable in SDN: policy envelopes bound
exploration and execution, action logs make local decisions auditable, and controller
arbitration resolves conflicts on shared bottlenecks.

Evaluation on a 1,024-host cloud testbed shows 35.5\% higher average core-link utilization
than Static ECMP, 40.3\% lower P99 elephant FCT, and 62.6\% fewer SLA violations compared
to Static Meter, with under 2.1\% per-agent CPU overhead and 12\,MB memory footprint.
Multi-agent coordination eliminates reroute oscillation with negligible additional control
traffic.

The main lesson is that locality alone is not enough: local learners become deployable in
SDN only when their authority is explicitly bounded, versioned, and auditable. Future work
includes hardware-fabric validation with P4 or SmartNIC feedback, stronger online-adapting
MARL and online-RL baselines, extensions to multi-domain and inter-AS traffic engineering,
and stronger multi-agent analysis for discrete rerouting and queue-priority actions.

\section*{Acknowledgment}

This work was supported by the New Generation Artificial Intelligence National Science and
Technology Major Project (No. 2025ZD0122304) and the Shenzhen Science and Technology
Program (Nos. ZDCY20250901095900001 and SYSPG20241211173440003).

\bibliographystyle{IEEEtran}
\bibliography{refs}

\end{document}